# Spatio-temporal optical vortices

N. Jhajj, I. Larkin, E.W. Rosenthal, S. Zahedpour, J.K. Wahlstrand, and H.M. Milchberg

*Institute for Research in Electronics and Applied Physics*
*University of Maryland, College Park, MD 20742*

**Abstract**

We present the first experimental evidence, supported by theory and simulation, of spatio-temporal optical vortices (STOVs). Quantized STOVs are a fundamental element of the nonlinear collapse and subsequent propagation of short optical pulses in material media. A STOV consists of a ring-shaped null in the electromagnetic field about which the phase is spiral, forming a dynamic torus which is concentric with and tracks the propagating pulse. Depending on the sign of the material dispersion, the local electromagnetic energy flow is saddle or spiral about the STOV. STOVs are born and evolve conserving topological charge; they can be simultaneously created in pairs with opposite windings, or generated from a point null. Our results, here obtained for optical pulse collapse and filamentation in air, are generalizable to a broad class of nonlinearly propagating waves.

**Introduction**

Vortices – localized regions in which the flow of some quantity such as mass or electromagnetic energy circulates about a local axis – are a common and fundamental element of classical (1) and quantum fluids (2,3,4) as well as optics (5,6). In quantum fluids, the circulating quantity is the spatial atomic probability density; in optics it is the electromagnetic energy density. Both densities are expressed as the magnitude squared of a complex scalar field $\psi = ue^{i\Phi}$ derivable from a Schrodinger-like equation (SE), where the vortex circulation is $\Gamma = \oint_c \nabla\Phi \cdot d\mathbf{l}$, and where $u$ and $\Phi$ are real scalar fields. A non-zero value of $\Gamma$ implies a discontinuity or 'defect' in $\Phi$. Furthermore, the single-valuedness of fields demands that the circulation be quantized: $\Gamma = 2m\pi$, where $m$, an integer, is called the 'topological charge'. Because $\Gamma$ remains constant as one shrinks the contour around the vortex core, the phase $\Phi$ becomes undefined at the core centre ('phase singularity') and the field magnitude is necessarily zero there.

In optics, vortices have been heavily investigated for decades, but as far as we know, the types that have been studied experimentally are entirely those that can be supported by monochromatic waves and are able to exist purely in the spatial domain. We note that there is a significant body of theory and simulation work dedicated to the study of spatio-temporal solitons (6), including those with embedded vorticity (7, 8).

Here, we present experimental evidence, backed by theory and simulation, for a new type of dynamic phase vortex that is embedded in the *spatio-temporal* optical field. The vortex takes the form of a torus that loops around the pulse propagation axis, and moves with the pulse. The spatiotemporal character of the vortex dictates that energy flow near the core is saddle or spiral depending on the sign of dispersion of the propagation medium. Further, we show that vortex formation is an inherent feature of nonlinear collapse arrest, and that vortex dynamics are associated with changes to the morphology of a nonlinearly propagating pulse.

In an early, influential publication, Nye and Berry introduced the concept of dislocations (field nulls) to wave theory and gave many examples (9). As a well-known example of a vortex in linear optics, a linearly polarized Laguerre Gaussian (LG) mode $E(r,\varphi,z)$ of non-zero integer order $l$ with azimuthal dependence $\exp(il\varphi)$ has an on-axis null in the field, a topological charge $l = \Gamma/2\pi$, an orbital angular momentum (OAM) of $Nl\hbar$, where $N$ is the number of photons in the beam, and a spiral energy density flux about its axis (10). Such modes can be generated by phase plates (11), and the phase singularity at beam centre is called a screw dislocation (9). LG modes are solutions to the paraxial wave equation $i\partial\psi/\partial z + \nabla_\perp^2 \psi = 0$ for propagation along $z$, which is of the SE form with zero potential (10). An example of an LG field with a null on axis is $E_{0,\pm1}(\mathbf{r}_\perp,z) \propto [x \pm iy] w_r^{-1}(z) \left\{ e^{-|\mathbf{r}_\perp|^2/w_r^2(z)} e^{i\Phi_p(\mathbf{r}_\perp,z)} \right\}$, where $\mathbf{r}_\perp = x\hat{\mathbf{x}} + y\hat{\mathbf{y}}$, $\Phi_p(\mathbf{r}_\perp,z)$ is the propagation phase, and $w_r(z)$ is the $z$-dependent spotsize.

Now consider, as an example, a construction similar mathematically to the LG field but physically quite different: the pulsed Gaussian field $E(\mathbf{r}_\perp,z,\xi) \propto [x/w_r(z) \pm i\xi/w_\xi(z)] \left\{ e^{-|\mathbf{r}_\perp|^2/w_r^2(z)} e^{-\xi^2/w_\xi^2(z)} e^{i\Phi_p(\mathbf{r}_\perp,z,\xi)} \right\}$, where $\xi = v_g t - z$ is the local position in a frame moving with the pulse group velocity $v_g$, and $w_\xi$ is the $z$-dependent axial pulse length. This field has a moving null at $(x=0, \xi=0)$ along the y-axis, perpendicular to the direction of propagation, and the circulation $\Gamma$ on closed spatio-temporal contours in $(x, \xi)$ around the null is quantized. We refer to this apparently previously unexplored object as a *spatio-temporal optical vortex* (STOV).

In this paper, we demonstrate that STOVs are a fundamental and universal feature of optical pulse collapse and arrest in self-focusing media. Their existence in nonlinear ultrafast pulse propagation appears to be ubiquitous, and their creation, motion, and destruction is strongly linked to the complex spatiotemporal evolution of the pulse.

**Spatio-temporal optical vortices (STOVs)**

Optical collapse is a fundamental phenomenon in nonlinear optics (12). It occurs when the laser pulse-induced change in the medium's refractive index generates a self-lens whose focusing strength increases with intensity. Above a critical power level ($P_{cr}$), self-lensing exceeds diffraction, and the pulse experiences runaway self-focusing. In the absence of "arrest" mechanisms terminating self-focusing, the pulse would collapse to a singularity. In reality, additional physical effects intervene. For example, in the case of femtosecond filamentation in ionizing media (13), plasma generation acts to defocus the pulse when the peak self-focused beam intensity reaches the ionization threshold. Other collapse arrest mechanisms include dispersion-induced pulse lengthening (14), cascaded third order nonlinearities (15), vectorial effects from beam non-paraxiality (16), and, in the case of relativistic self-focusing, electron cavitation (17).

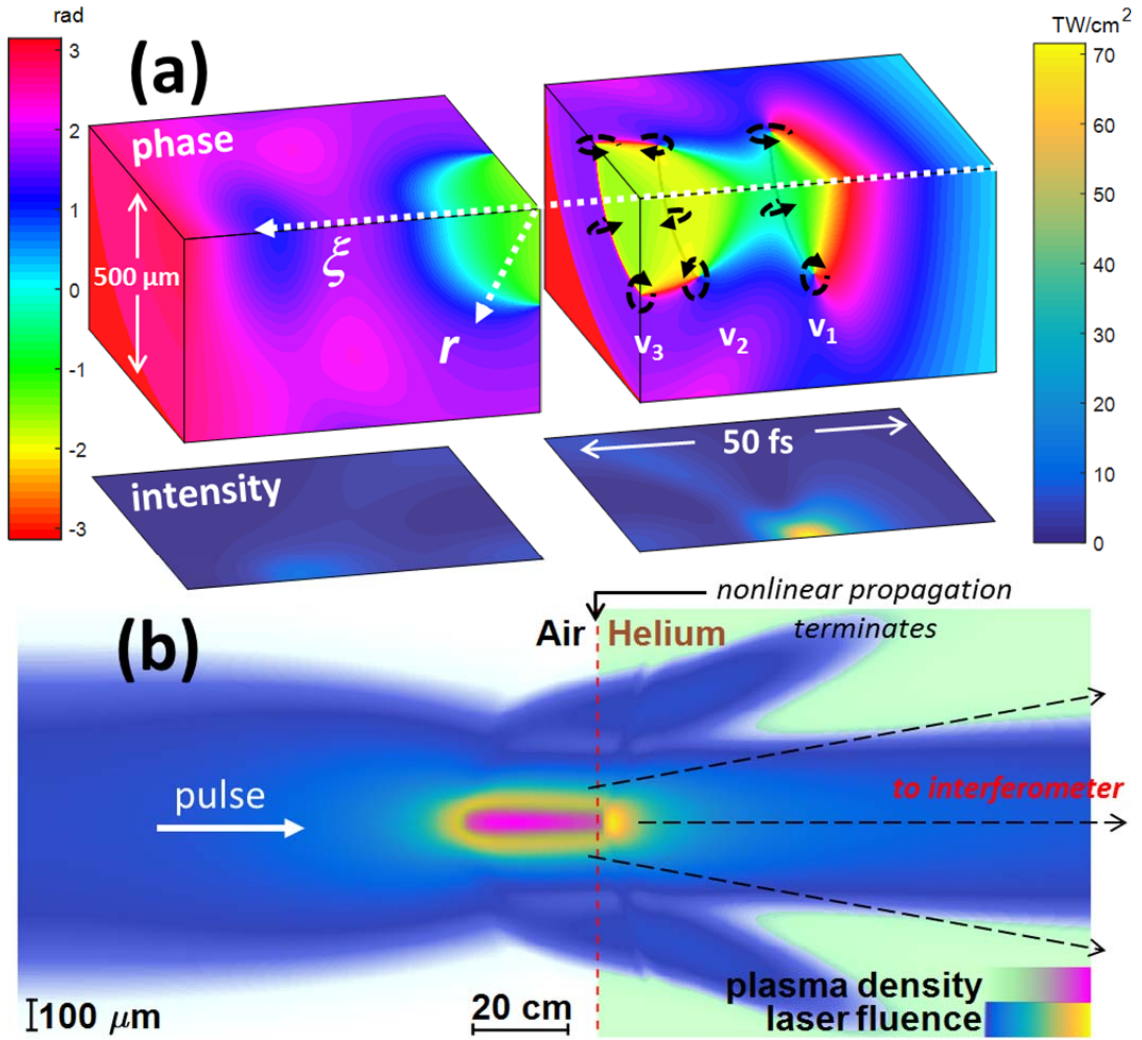

**Figure 1.** (a) Phase and intensity projections of simulated pulse, in local coordinates, showing STOV generation. (b) Simulation of an air filament crossing the air-helium boundary for the conditions of (a), showing the beam fluence and plasma density. Nonlinear propagation terminates as the beam transitions from air to helium, whereupon the beam and a reference pulse (not shown, see (24)) are directed to an interferometer.

In both calculations and simulations (18,19), it has been found that the following modified paraxial equation for wave evolution, in the form of a time-dependent nonlinear Schrödinger equation (NLSE), is well-suited to describe optical pulse collapse and collapse arrest for beam propagation along $z$:

$$2\frac{\partial}{\partial z}\left(ik-\frac{\partial}{\partial \xi}\right)\psi+\nabla_\perp^2\psi-\beta_2\frac{\partial^2\psi}{\partial \xi^2}+k^2V\{\psi\}=0 \qquad (1)$$

Here, $E=\psi(\mathbf{r}_\perp,z,\xi)e^{i(kz-\omega t)}$ is the dimensionless electric field component, $\psi$ is the pulse envelope, $\beta_2=c^2k_0(\partial^2 k/\partial\omega^2)_0$ is the dimensionless group velocity dispersion (GVD) at $\omega=\omega_0$, $\mathbf{r}_\perp$ and $\xi$ are as before, and the axial position $z$ can be viewed as a time-like coordinate. The physics of self-focusing and arrest is contained in the functional $V\{\psi\}$. To demonstrate STOV generation in the arrest of self-focusing collapse, we perform simulations using Eq. (1), assuming azimuthal symmetry, and include electronic, rotational, and ionization nonlinearities in $V\{\psi\}$ (18,20). The Gaussian input pulse is 3 mJ, 45 fs (Gaussian FWHM), with input waist $w_0$=1mm. Figure 1(a) is a post-collapse plot of the pulse phase at $z$=120 cm, which shows the emergence of two oppositely wound and oppositely propagating STOVs ($v_1$ (+1 charge, forward) and $v_2$ (-1 charge, backward)) entrained between the higher intensity core of the beam and the beam periphery. The delayed rotational nonlinearity from the $N_2$ and $O_2$ air constituents (21) forms an additional vortex pair ~100 fs behind the main pulse, $v_3$ (+1, forward) and $v_4$ (-1, backward), where $v_3$ is shown bisected in Fig. 1(a) and $v_4$ has exited the simulation window. Further evolution of the simulation shows that $v_2$ and $v_3$ collide and annihilate, while $v_1$ continues propagating with the most intense part of the pulse (to be discussed later). We note that the delayed generation of $v_3$ and $v_4$, where the pulse intensity is many orders of magnitude weaker than at its peak, shows that STOVs can also be generated *linearly* by an imposed spatio-temporal index transient. STOVs are not merely mathematical free-riders on intense propagating pulses: as we will see, a real energy flux **j** circulates either as a saddle (for $\beta_2$>0) or as a spiral (for $\beta_2$<0) in the ($\mathbf{r}_\perp,\xi$) plane surrounding the STOV core.

To understand the generation of these spatio-temporal toroidal structures, recall that phase vortices in fields are closely associated with localized field nulls (9). In arrested self-focusing, field nulls occur as a natural part of the dynamics and spawn toroidal vortices of opposite charge. This can be illustrated by the toy model of Fig 2, which shows the effect on a beam of abruptly spatially varying self-induced change in refractive index, which we model here as a sharp transverse step. We consider right-to-left propagation of the 'half plane-wave' pulse $E(x,z,\xi)=E_0(\delta+H(x))e^{-(\xi/\xi_0)^2}e^{i(k(z-\omega/v_g)-\omega\xi/v_g)}e^{i\phi_{NL}(\xi,x,z)}$, neglecting dispersion and diffraction, where $\xi$ and $z$ are as before, $\delta$<<1, $H(x)$=1 for $x$>0 and 0 otherwise, $\phi_{NL}(x,z,\xi)=kn_2|E(x,z,\xi)|^2 z$ is the sharply stepped nonlinear Kerr phase, and $n_2$ is the nonlinear index of refraction. At $z$=0, the phase fronts are aligned for all $x$. As the pulse propagates, the Gaussian intensity distribution causes the front of the pulse to red shift and the back of the pulse to blue shift, with spatio-temporal phase front shear developing between the pulse 'core' at $x$>0 and periphery at $x$<0. When the propagation reaches $z=z_v=\pi(kn_2E_0^2)^{-1}$, the peak of the pulse in the core is $\pi$ out of phase with the periphery, and the electric field magnitude nulls out at the single point $(\xi=0, x=0, z=z_v)$, marked by a circle, forming an 'edge dislocation' (9). Upon further propagation, continued phase shearing spawns two null points of opposing (±1) phase winding, marked by circles ($z>z_v$). These two STOVs, whose axes are along $y$ (perpendicular to page), immediately begin moving apart; one vortex advances in time towards the front of the pulse while the other vortex moves to the back. Similar dynamics are theorized to exist in monochromatic breather-solitons, where ring shaped vortices are formed quasi-periodically throughout propagation (22).

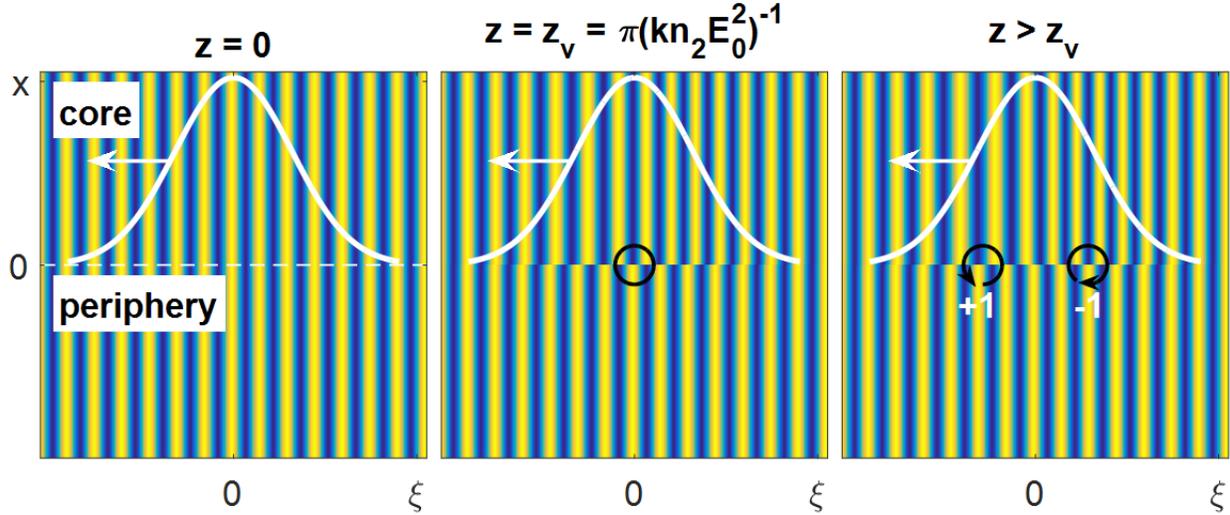

**Figure 2.** Toy model showing birth of a vortex pair via spatio-temporal phase front shear. The white curve and arrow depicts the axial (temporal) intensity profile and propagation direction, while the "core" and "periphery" labels denote the spatial intensity step. The $z=0$ panel shows the initial condition where phases are aligned, the $z=z_v$ panel shows the birth of the null (vortices overlap) and the $z>z_v$ panel shows continued shear carrying the vortices apart, with the +1 vortex moving to the temporal front, and the -1 vortex moving backward.

These general features occur in simulations of self-focusing collapse arrest. In Fig. 3, we show 2D profiles of the pulse's phase $\phi(r,z,\xi=\xi_v)$ and intensity $c|E(r,z,\xi=\xi_v)|^2/8\pi$ for a simulation using Eq. (1), tracking the moving plane $\xi=\xi_v$ where the phase singularity and field null first appear. Log lineouts of the intensity (normalized to $10^{13}$ W/cm$^2$) are overlaid on the 2D profiles. As the initial ($z=0$) Gaussian input beam self-focuses, a strongly peaked high intensity central core (similar to the Townes profile (23)) develops, surrounded by a lower intensity periphery, with a sharp transition knee between them ($z=156$ cm). The associated phase plot shows the central core having accumulated a much larger nonlinear phase shift than the periphery. During collapse, the knee moves radially inward, preventing phase shear from accumulating substantially at any particular radial location. However, as the core peak intensity rises to the point where ionization begins ($z=160$ cm, not shown), the location of the intensity knee stabilizes and highly localized phase shear builds up, greatly steepening the transition between core and periphery, with the field beginning to dip towards a null ($z=166$cm). Finally, only a short propagation distance later ($z=167$cm), sufficient shear develops that the core and periphery are $\pi$ out of phase at the slice $\xi=\xi_v$, with the dip in the simulation becoming orders of magnitude deeper, forming a ring-shaped null surrounding a core of relatively flat intensity and phase. Note the core-periphery phase difference $\Delta\phi_{cp}$ jumps by $2\pi$ between $z=166$ and $167$ cm. (As seen in the computation of Fig 1(a), the null then spawns two oppositely charged ($\pm 1$) toroidal vortices that propagate forward and backward.) The plane $\xi=\xi_v$ is still shown at $z=179$cm, but the vortices have now migrated out of that plane. Later in the paper, we derive and discuss equations of motion describing vortex dynamics within the moving reference frame.

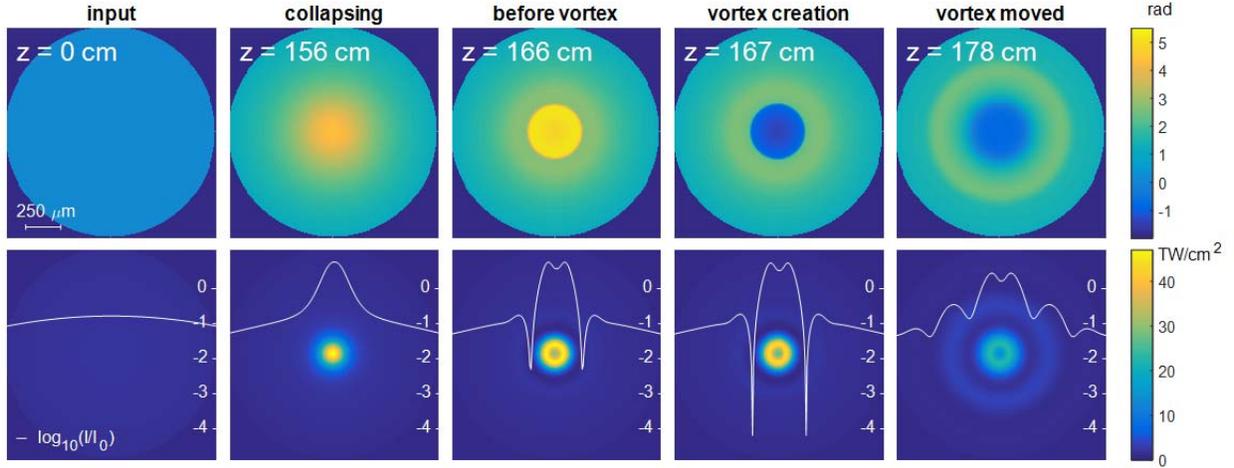

**Figure 3.** Simulations of beam phase (top) and intensity (bottom) at the axial/temporal slice $\xi_v$ where the STOV pair first appears. From left to right, the pulse is advancing along z, with the input shown at z=0, a collapsing beam at z= 156cm, ionization onset at 160 cm (not shown), just before the vortices spawn at 166cm, just after the vortices spawn at 167 cm, and an image where the vortex pair has moved out of the $\xi=\xi_v$ plane. The linear intensity images are overlaid with centered lineouts of $\log_{10}(I/I_0)$, where $I_0=10^{13}$W/cm$^2$. Experimental parameters were used as code inputs: $w_0 = 1.3$mm, pulse energy = 2.8mJ ($P/P_{cr} = 6.4$), pulse FWHM 45fs.

**Experiment**

In order to experimentally confirm the existence of STOVs, we image the *spatio-spectral* phase and intensity profiles of femtosecond laser pulses mid-flight during their pre- and post-collapse evolution in air. Until now, we have been discussing STOVs as a spatio-temporal phenomenon, but they are also vortices in their spatio-spectral representation. This has enabled us to unambiguously observe them (24).

Direct measurement of the spatial phase and intensity profile of a filament in mid-flight is not amenable to standard techniques; the use of relay imaging or beam splitters is subject to the severe distorting effects of nonlinear propagation and material damage. However, by interrupting nonlinear beam propagation by an air-helium interface, the in-flight beam intensity and phase profile can be *linearly* imaged through helium, taking advantage of the very large difference in instantaneous nonlinear response between helium and air ($n_{2,He}/n_{2,air} \approx 0.04$ (21,25)). This technique was first employed by Ting *et al.* (26) to measure the in-flight intensity profile of a femtosecond filament. Here we extend the technique to also enable measurement of the pulse transverse phase profile. A simple view of the scheme is shown in Fig. 1(b), where we show the post-collapse pulse simulated in Fig. 1(a) encountering an air-helium interface and terminating nonlinear propagation, after which it is imaged to the output of a Mach-Zehnder interferometer and combined there with a weak femtosecond reference beam. A detailed description of the experimental setup is found in the Supplementary Material (24).

Figure 4 shows the beam on-axis phase shift $\Delta\phi$ with respect to the interferometric reference pulse as a function of $P/P_{cr}$ at a fixed position of $z_h$ =150cm after launch, where $P_{cr} = 3.77\lambda^2/8\pi n_0 n_2$ for our Gaussian input beam profile and $P = \varepsilon_i / \tau$ is the input power. The red points are averages over 2600 shots (blue points) in 150 energy bins. It is important to note that the scatter in $\Delta\phi$ of roughly +/- 1 rad at any given laser power is constant across all powers measured, including $P/P_{cr} \ll 1$, where we could not detect any nonlinear phase. Therefore, the scatter is due to the shot-to-shot interferometric instability of the measurement and is not intrinsic to the filamentation process. The most striking aspect of the plot is the abrupt jump in beam central phase of approximately $\sim 2\pi$ at $P/P_{cr} \sim 5$. The phase goes from positive and rapidly increasing (increasing self-focusing) to abruptly negative (defocusing), providing a clear signature of the transition from the pre-collapse to post-collapse beam. For nominally constant laser power right at the jump, the phase fluctuates in the range $\sim \pm \pi$, showing the extreme sensitivity of the phase flip.

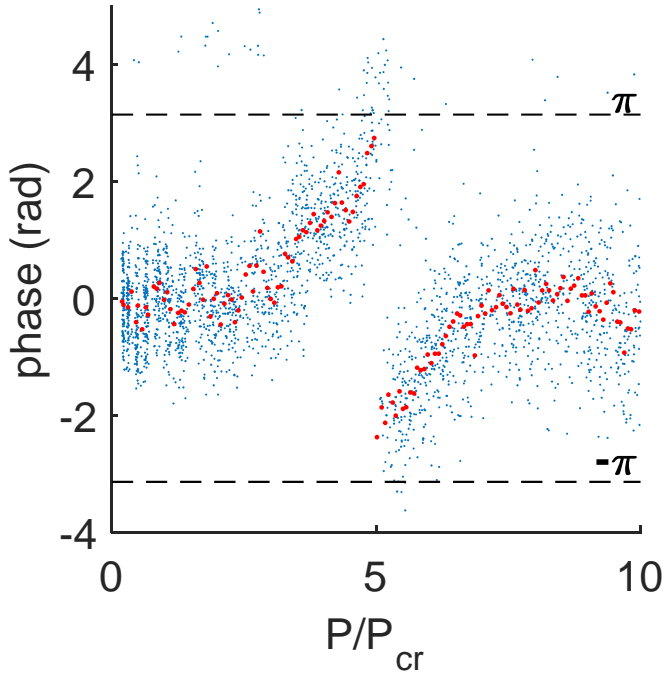

**Figure 4.** Beam on-axis phase shift (with respect to flat-phase reference arm) as a function of pulse power at $z_h$ = 150 cm. The phase jumps abruptly by $\sim 2\pi$ around $P/P_{cr}$ = 5, providing a clear signature of the transition from the pre-collapse to post-collapse beam. The red points are averages over 2600 shots (blue points) in 150 energy bins.

A more revealing way to display what is happening at the collapse is shown in Fig. 5. Here, for given $z_h$, the phase images $(z_h, \varepsilon_i)$ were searched for $\varepsilon_i$ or $P/P_{cr}$ where the central phase appeared to randomly flip sign from shot to shot. These are the powers at which pulse collapse was observed for each position, just as $P/P_{cr} \sim 5$ was for $z_h=150$cm in Fig. 4. The top row of Fig. 5 shows beam phase and intensity images for input power $P/P_{cr}=4.4$ (at $z_h= 165$ cm). Because the onset of collapse arrest is extremely sensitive to fluctuations in the beam energy (as seen in Fig. 4), these images span the possibilities of pre-arrest through post- arrest of the collapse, and typically three types of images appear. Panel (i) shows strongly-peaked intensity and positive phase; the beam is collapsing, but arrest has not yet begun. Panel (ii) shows radically different images, as does panel (iii): the intensity images show narrow ring-shaped nulls embedded in a relatively smooth background, and the phase images show a sharp yet smoothly transitioning jump close to $\pi$ or $-\pi$ across the rings, with the phase jumps flipped between (ii) and (iii). We note that the smooth phase transition from the periphery to the core rules out $2\pi$ phase ambiguities in interferometric phase extraction.

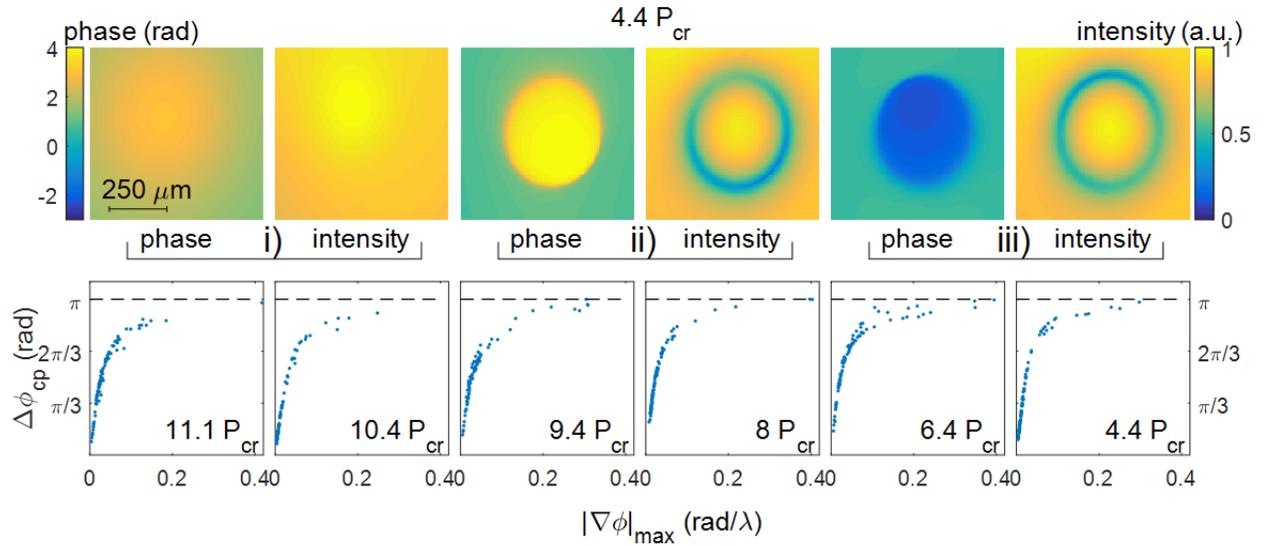

**Figure 5.** Top row: retrieved spatial phase and intensity images at z = 165cm, $P/P_{cr}$ = 4.4 for i) a pre-collapse beam, ii) and iii) beams where a vortex ring is on either side of the reference central wavelength of 800 nm. The bottom row shows that as the maximal phase gradient in the images increases, the maximal phase shift saturates at π for all cases of $P/P_{cr}$ leading to beam collapse.

The bottom row of Fig. 5 plots, for a range of $P/P_{cr}$, the phase difference $|\Delta\phi_{cp}|$ between the core and periphery of the beam. To do this, for each phase image we compute the difference between the maximum and minimum values of the phase within a 60 micron box centered about the largest spatial phase gradient, the radial location of which defines 'core' and 'periphery'. For each nominal value of $P/P_{cr}$, it is clear that as the phase gradient becomes large, the phase difference saturates at π. Near saturation, roughly 50% of the shots have the core phase advanced from the periphery while the others show the reverse.

The evidence from Figs. 4 and 5, and comparison to the simulation of Fig. 3, strongly suggest that we are imaging spatio-temporal vortex rings: the abrupt appearance of ring-shaped nulls in the field magnitude accompanied by 2π phase jumps in $|\Delta\phi_{cp}|$ across the null– these are exactly the signatures of a vortex. Because the circulation around a general singly charged vortex is 2π, examining our vortex in the spatio-spectral domain $(\mathbf{r}_\perp, \omega_\xi)$, one would expect, depending on the sign of vortex winding, that the core-periphery phase difference $|\Delta\phi_{cp}|$ jumps by 2π from $\omega_\xi$ slices just before the vortex ($\Delta\phi_{cp} = \pm\pi$) to $\omega_\xi$ slices just after ($\Delta\phi_{cp} = \mp\pi$). For example, before a vortex of charge +1 the core is phase-advanced with respect to the periphery; after the vortex it is phase-retarded. This is exactly what we observe experimentally and what is predicted in the simulation of Fig. 3 and its spatio-spectral counterpart, where even the ~400μm diameter of the vortex ring is accurately determined. Of the four STOVs simulations show are generated at collapse arrest in air (see discussion of Fig. 1(a)), only the temporally foremost +1 STOV does not annihilate or separate from the bulk of the pulse. Using Fig. 1(a) as a guide, we interpret our results as a spectral "fly-by" of a +1 STOV from the blue to the red side of our reference pulse spectrum centered at 800nm. We note that a similar fly-by of a −1 STOV from red to blue would present itself in an identical manner.

How are STOVs born in real collapsing pulses? In our $(r, z, \xi)$ simulations, vortices are immediately born as tori around the beam owing to azimuthally symmetric (φ-independent) phase shear. In real beam collapse, where there is $\varphi$ variation in the laser field, topological considerations lead us to expect that shear in higher E-field locations will first lead to a point null, followed by an expanding and distorted torus on one side of the beam that progressively wraps to the other side of the beam and then, meeting itself, splits into two toroidal STOVs of opposite phase winding. The onset of these STOVs, aligned with planes of constant $\xi$, has a beam-regularizing influence, as seen in the images of Fig. 5, which show remarkably flat phase and intensity profiles inside the ring. This could be the reason for the notably high quality spatial modes and supercontinuum beams (so-called 'spatial cleaning' (27)) observed in filamentation. We are performing 3D propagation simulations to verify this scenario. We also note that the ring null forms a natural and well-defined boundary between what had been qualitatively labelled the 'core' and 'reservoir' regions (13) in femtosecond filaments.

**STOV dynamics and energy flow**

Once STOVs are generated, it is important to understand how they propagate. Following the method of ref. (28) as applied to OAM vortices, we approximate the local form of the STOV as a spacetime "R-vortex" $\psi_{vortex} \equiv (\xi - \xi_0) \pm i(r - r_0)$ of charge ±1 with a linear phase winding about $(\xi_0, r_0)$, embedded in a background field envelope $\psi_{bg}$ such that $\psi = \psi_{bg}\psi_{vortex}$. If we take $\psi_{bg} = \rho e^{i\chi}$, where $\rho$ and $\chi$ are the real amplitude and phase of the background field, then as the pulse propagates along $z$, the NLSE (Eq. (1)) in $(r, z, \xi)$ moves the vortex location $\mathbf{r}_{vortex} = (r_0, \xi_0)$ according to (24):

$$\left[ k \frac{\partial \mathbf{r}_{vortex}}{\partial z} = \pm \frac{1}{\rho}\left( \hat{\mathbf{r}} \frac{\partial \rho}{\partial r} - \hat{\xi}\beta_2 \frac{\partial \rho}{\partial \xi} \right) \times \hat{\boldsymbol{\varphi}} + \left( \hat{\mathbf{r}} \frac{\partial \chi}{\partial r} - \hat{\xi}\beta_2 \frac{\partial \chi}{\partial \xi} \right) \pm \hat{\xi} \frac{1}{2r} \right]\bigg|_{(r,\xi,z)=(r_0,\xi_0,z_0)}, \quad (2)$$

where the derivatives are evaluated at the present vortex core location $(r_0, \xi_0, z_0)$. Equation (2) demonstrates interesting analogies with fluid vortices. The term $\pm \hat{\xi}/2r$ propels the vortex forward or backward depending on its charge and radius (curvature), and strongly resembles the speed $\sim \Gamma/4\pi r$ of a toroidal fluid vortex (such as a smoke ring) (29). Identifying $\hat{\mathbf{r}} \partial \chi/\partial r = \nabla_\perp \chi = k \mathbf{j}_\perp / \rho^2$ as the local effective velocity associated with the background electromagnetic flux (24), we interpret it as a charge-independent drag-like term, expanding (contracting) the STOV for power outflow (inflow) for $\partial \chi/\partial r > 0 \, (< 0)$. The term $\pm \hat{\mathbf{r}} \times \hat{\boldsymbol{\varphi}} \, \rho^{-1} \partial \rho/\partial r$ is a Magnus-like motion (1), here propelling the STOV along $\pm \hat{\xi}$, perpendicular to the vortex circulation vector $\hat{\boldsymbol{\varphi}}$ and the ring expansion/contraction direction $\hat{\mathbf{r}}$. The terms $-\hat{\xi}\beta_2 \partial \chi/\partial \xi$ and $\mp \hat{\xi} \times \hat{\boldsymbol{\varphi}} \beta_2 \rho^{-1} \partial \rho/\partial \xi$ are their spatio-temporal analogues. In gases, small $\beta_2$ (~$10^{-5}$) makes these terms negligible; they contribute much more significantly in solid media.

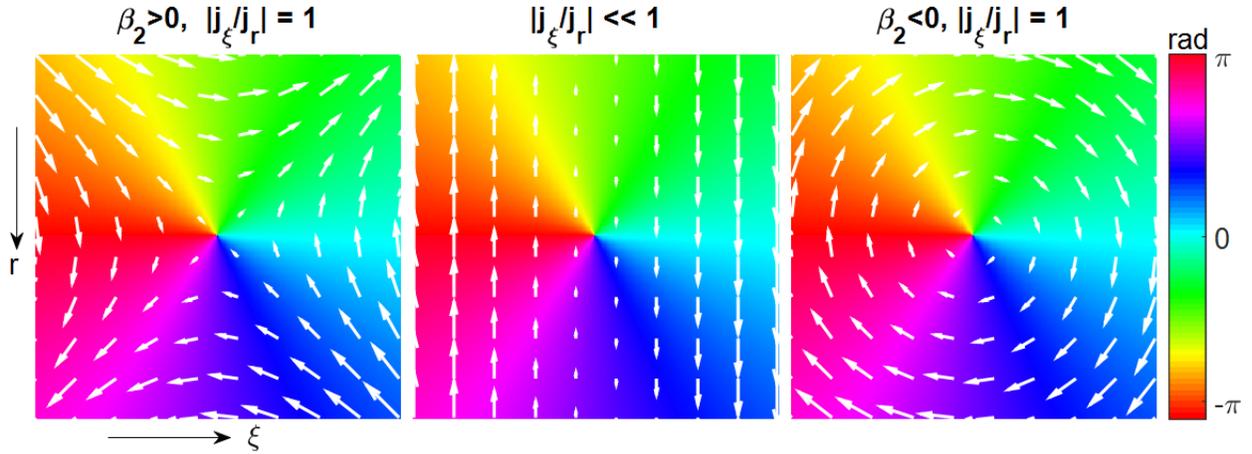

**Figure 6.** Demonstration of energy flow about an "$r$-vortex" STOV. White arrows correspond to size and direction of **j** in Eq. (3). There are three distinct regimes: the left panel shows the saddle regime which exists in regularly dispersive media ($\beta_2>0$), the middle panel shows the degenerate regime which can exist in regular or anomalously dispersive media and has a dominant axis for energy flow, and the right panel shows the spiral regime which exists in anomalously dispersive media ($\beta_2<0$).

We note that $\psi_{bg}$ is not a fixed field independent of vortex motion. Equation (2) should be understood as a stepwise predictor of vortex motion based on an updated $\psi_{bg}$. To understand how STOVs affect the background field, it is useful to consider the electromagnetic energy flux associated with the full field envelope $\psi = \psi_{bg}\psi_{vortex} = ue^{i\Phi}$ in the moving frame of the pulse (24),

$$\mathbf{j} = \frac{1}{k}u^2(\nabla_\perp \Phi - \beta_2 \frac{\partial \Phi}{\partial \xi}\hat{\xi}) \ , \tag{3}$$

where one can see that the sign of $\beta_2$ determines whether the energy flow near the STOV is spiral ($\beta_2 < 0$) or saddle ($\beta_2 > 0$). What are the relative contributions of longitudinal and transverse energy flow about a STOV? For filamentation in air, $\beta_2 \sim 10^{-5}$, the characteristic axial pulse length and width are $c \times 10$ fs and 100 μm (filament core), and one finds $j_\xi / j_r \sim 10^{-4} \ll 1$. Here, the distinction between saddle and spiral is not important, as we are in the degenerate case. For solids, however, where $\beta_2 \sim 10^{-2}$, we expect the distinction between saddle and spiral energy flow to be very important. Figure 6 provides intuitive visualization of the energy flow pattern for the saddle, degenerate and spiral cases.

Equations (2) and (3) are useful for an intuitive picture of the governing dynamics, especially when viewed together with propagation simulations. For example, for the four STOVs seen in the simulations of Fig. 1(a), their dominant early movement is governed by $\pm\hat{\xi}\rho^{-1}\partial\rho/\partial r$ in Eq. (2), which propels the +1 (-1) STOV temporally forward (backward), with the forward motion initially being superluminal. (We will explore the detailed implications of superluminal STOV motion in a future publication.) A consequence of the opposing directions for the ± charges is collision and annihilation of $v_2$ and $v_3$ of Fig 1(a), as discussed earlier. Remarkably, the $v_3$ STOV superluminally climbs from a region of negligible intensity through many orders of magnitude of increasing intensity to reach and annihilate $v_2$, whereupon a local depression is left in the field that more gradually dissipates. The $v_1$ STOV eventually settles in the temporal middle of the highest intensity portion of the pulse, propagating at nearly $v_g$. Evidently, $\rho$ self-consistently evolves to balance the $\hat{\xi}$ terms in Eq. (2) and $\chi$ flattens along the radial dimension (as indicated in the experiment), preventing expansion or contraction of the STOV. Our simulations show that a surviving +1 STOV is always coupled to the 'optical bullet' forming the most intense part of the filament. This is no coincidence, as the energy flow for a +1 STOV is toward (away from) the pulse axis temporally in front of (behind) the STOV. This is exactly as expected, where the front of the pulse draws energy in by Kerr self-focusing, and energy at the back of the pulse is directed outward by plasma refraction. A link to movies of STOV dynamics is available in Ref. (24).

In real beams without φ symmetry, we expect collisions of oppositely charged STOVs to be much more complex, although the beam regularization observed in experiments may conspire to promote collisions. Auxiliary 3D+1 linear propagation simulations in which we imposed STOVs as initial conditions on Gaussian beams show repulsion of like-charged STOVs, which pass around each other, and splitting of higher charge STOVs into multiple STOVs of single charge. We note that our measured air-based STOVs are not solitons, as diffraction does not balance self-focusing for a dark object. STOV solitons could exist, however, in an anomalously dispersive, self-defocusing medium (6).

**Conclusions**

We have introduced the concept of the spatio-temporal optical vortex (STOV) to ultrafast optics and demonstrated its existence. STOVs form naturally as a consequence of arrested self-focusing collapse and their dynamics influences subsequent pulse propagation. STOVs can also be imposed linearly via prescribed spatio-temporal or spatio-spectral phase shifts, making possible their engineering for applications. While evidence for STOV generation was demonstrated in experiments and simulations of short pulse filamentation in air, we expect that STOVs, whose dynamics are subject to topological constraints, are a fundamental and ubiquitous element of nonlinear propagation of intense pulses.

Substantial work remains to be done in studying the nonlinear propagation dynamics of beams with STOVs, where the size and sign of dispersion determines the nature of the local energy flow, creating three distinct regimes: saddle, spiral, and degenerate. STOV-STOV interactions should prove to be a fundamental mediator of intra- and inter-beam dynamics.

**Acknowledgements**

The authors thank John Palastro (NRL) for the use of his propagation simulation (and permission to modify the source code), and Tony Ting (NRL), Dan Lathrop (UMd), and Peter Megson (UMd) for useful discussions. This work is supported by the Defense Advanced Research Projects Agency (DARPA), the Air Force Office of Scientific Research (AFOSR), the National Science Foundation (NSF), and the Army Research Office (ARO).

**Supplementary Materials**

*See below*

**References and Notes**

1. G. K. Batchelor, *An Introduction to Fluid Dynamics*, Cambridge University Press (1977).

2. M. J. H. Ku *et al.*, Phys. Rev. Lett. **113**, 065301 (2014)

3. S. Donadello *et al.*, Phys. Rev. Lett. **113**, 065302 (2014)

4. G.P. Bewley, D. P. Lathrop, and K. R. Sreenivasan, Nature **441**, 588 (2006)

5 . M. R. Dennis, Y. S. Kivshar, M. S. Soskin and G. A. Swartzlander Jr., J. Opt. A: Pure Appl. Opt. **11**, 090201 (2009)

6 . B.A. Malomed, D. Mihalache, F. Wise and L. Torner, J. Opt. B **7,** R53 (2005)

7 . D. Mihalache et al., Phys. Rev. Lett. **97**, 073904 (2006)

8 . Y. V. Kartashov, B. A. Malomed, Y. Shnir, and L. Torner, Phys. Rev. Lett. **113**, 264101 (2014)

9 . J. F. Nye and M. V. Berry, Proc. R. Soc. Lond. A. **336**, 165 (1974).

10 . L. Allen, M. W. Beijersbergen, R. J. C. Spreeuw, and J. P. Woerdman, Phys. Rev. A **45**, 8185 (1992).

11. M. W. Beijersbergen *et al*. Optics Communications **112**, 321 (1994).

12. G. Fibich, *The Nonlinear Schrödinger Equation: Singular Solutions and Optical Collapse*, Springer (2015)

13. A. Couairon and A. Mysyrowicz, Phys. Rep. **441**, 47 (2007)

14. D. Strickland and P. B. Corkum, J. Opt. Soc. Am. B **11**, 492 (1994)

15. P. Panagiotopoulos, P. Whalen, M. Kolesik, and J. V. Moloney, Nat. Photonics **9**, 543 (2015).

16. G. Fibich and B. Ilan, Physica D **157**, 112 (2001)

17. A.J. Goers *et al*., Phys. Rev. Lett. **115**, 194802 (2015)

18. J. P. Palastro, T. M. Antonsen Jr., and H. M. Milchberg, Phys. Rev. A **86**, 033834 (2012)

19. M. Kolesik and J. V. Moloney, Phys. Rev. E 70, 036604 (2004); J. Andreasen and M. Kolesik, Phys. Rev. E **86**, 036706 (2012)

20. E W Rosenthal *et al*., J. Phys. B **48**, 094011 (2015)

21. J. K. Wahlstrand, Y.-H. Cheng, and H. M. Milchberg, Phys. Rev. A **85**, 043820 (2012)

22 . A. S. Desyatnikov *et al*., Sci. Reports **2** , 771 (2012)

23. K. D. Moll, A. L. Gaeta, and G. Fibich, Phys. Rev. Lett. **90,** 203902 (2003)

24. See Supplementary Materials.

25. J. K. Wahlstrand, Y.-H. Cheng, and H. M. Milchberg, Phys. Rev. Lett. **109**, 113904 (2012)

26. A. Ting *et al*., Applied Optics **44**, 1474 (2005)

27. B. Prade *et al*., Opt. Lett. **31**, 2601 (2006)

28. D. Rozas, C. T. Law, and G. A. Swartzlander, Jr., J. Opt. Soc. Am. B **14**, 3054 (1997)

29. I. S. Sullivan *et al*., J. Fluid Mech. **609,** 319 (2008)

# Supplementary Materials

Experimental apparatus

The functional aim of the experimental apparatus is reconstruction of the spatial phase and intensity of a femtosecond optical air filament in mid-flight. To do this, we use an abrupt air-helium transition to halt nonlinear propagation, as ionization yield and self-focusing are both negligible in helium. After the filamenting beam is allowed some distance to diffract within the helium medium, it can be attenuated by wedges and the phase and intensity can be reconstructed using standard interferometric techniques (1,2).

The experimental setup is shown in Fig. S1. Our filamentation source is a chirped pulse amplification Ti:Sapphire amplifier ($\lambda$ = 800 nm, 45 fs, 0-5 mJ). The beam from the laser is spatially filtered using a pinhole to produce a Gaussian mode with flat phase fronts – this is important for the reference arm in the experiment, which requires a flat spatial phase. After spatially filtering, the filamenting and reference arms are generated using uncoated wedges in a Mach-Zehnder (MZ) configuration to create a large difference in power between the two beams (~$10^4$ : 1). Here, the low power reference arm reflects off the front faces of the wedges, while the high power signal arm transmits, creating a dispersion imbalance that is corrected further downstream. The beams are then compressed, with the signal arm (or pump/filamenting arm) now at 45 fs full width at half maximum (FWHM) intensity, rotated 90 degrees using a periscope (converting P-polarization to S), and down collimated to a waist of 1.3 mm using a reflective off-axis telescope.

After down-collimation, the beams are launched a variable distance spanning 50-225 cm beyond the last optic of the telescope before nonlinear propagation terminates inside the nozzle of the translatable helium cell (described in the next section). Past the air-helium transition, both beams propagate linearly in helium 50 cm, with the intense filament core of the high power beam expanding transversely in size. Both beams are then wedge-attenuated before being directed out of the cell through a 200 µm thick BK7 window. Outside the cell, the high power beam is attenuated to match the power of the reference arm via reflections from a second set of wedges in MZ configuration. In order to maintain polarization purity, polarization rotation from the upstream periscope was necessary, as wedge reflections preferentially select for s-polarization. Wedge transmissions by the reference arm fix the dispersion mismatch created in the pre-compressor MZ interferometer. The beams are recombined at the output of the interferometer and sent through a lens which images an upstream plane, just before the nozzle's gas transition region, to a CCD camera.

Air-Helium interface

The air-helium interface is formed by the non-turbulent flow of helium, at slightly positive pressure, into the ambient air through a ¼" diameter nozzle on a translatable rail-mounted cell. The filament propagates from air into the nozzle and nonlinear propagation terminates over the sharp 4 mm transition from air to helium. The helium-air transition was measured, as in Ting *et al.* (3), by monitoring the strength of a helium line at $\lambda$ = 588 nm as the helium cell nozzle is moved through a tightly focused 800 nm, 45 fs ionizing beam. As seen in Fig. S2a, the rapid drop-off of the helium line indicates that there is negligible helium beyond a 4 mm 10%-to-90% transition layer at the nozzle.

To confirm the fidelity of imaging and phase reconstruction through the helium cell, we used a time domain propagation code (4) in 2+1 dimensions ($r$, $z$, $\xi$) to model the propagation of a filamenting beam through the 4 mm air-helium transition into the far field in the bulk helium at atmospheric pressure. The accuracy of phase and intensity reconstruction was verified by reverse-propagating the beam via phase conjugation through vacuum back to the air region just before the transition. The results are displayed in Fig. S2b and S2c, which show that the reconstructed spatial intensity and phase at 800 nm (red) closely track that of the input electric field just before the transition region (black). In addition, we verified that small deviations from the correct imaging plane (+/- 1 cm) did not affect the results.

Interferometric reconstruction

Since the collimated low power reference arm has a flat spatial phase, the spatial phase difference, extracted from the interferogram using standard techniques (1,2), is just the spatial phase accumulated by the nonlinearly propagating beam. As such a pulse can develop a complicated time (and therefore frequency) dependence (5, 6), we note that what is actually measured is a weighted average of the spatial phase as a function of frequency, with the weight given by a product of the spectral amplitude of the low power reference arm and the spectral amplitude of the filamenting arm. The oscillatory portion of the signal on the CCD is given by

$$\text{int}(x,y) = 2\,\text{Re}\left[ e^{-ikx\sin\theta} \int_{-\infty}^{\infty} d\omega\, A_{ref}^{*}(x,y,\omega) A_{sig}(x,y,\omega) \right],$$

where $x$ and $y$ are transverse coordinates in the beam, $A_{ref}$ and $A_{sig}$ are Fourier transforms of the field envelopes of the reference and signal pulses, $k$ is the central wavenumber and $\theta$ is the crossing angle of the two beams. As the spatially filtered reference pulse propagates linearly, $A_{ref}$ is fully known, and is well-approximated by a Gaussian with flat spectral phase, $A_{ref}(x,y,\omega) = A_0 e^{-(x^2+y^2)/w^2 - \omega^2/\omega_0^2}$, and the weighted spatial-spectral phase of the signal beam is then extracted (1, 2). We simulated the interferometric results using propagation code (4) outputs, and found that in practice, the weighted spatial phase closely resembled the spatial phase at the central wavelength 800 nm, even when considering the sudden phase flips associated with the vortex.

Spatio-temporal optical vortex equations of motion

We assume azimuthal symmetry and consider the evolution of the complex envelope $\psi$ associated with the scalar electric field $E$,

$$E(r,\xi,z) = \psi(r,\xi,z) e^{i(kz-\omega t)}, \tag{S1}$$

where $k = 2\pi/\lambda$, $\xi = v_g t - z$, and $v_g$ is the group velocity. In the paraxial and slowly-varying envelope approximation, the propagation equation for $\psi$ is

$$2ik\frac{\partial \psi}{\partial z} + \nabla_\perp^2 \psi - \beta_2 \frac{\partial^2 \psi}{\partial \xi^2} + k^2 V(\psi)\psi = 0, \tag{S2}$$

Where $\nabla_\perp^2 = \partial^2/\partial r^2 + (1/r)(\partial/\partial r)$, $\beta_2 = c^2 k_0 (\partial^2 k/\partial \omega^2)_0$ is the dimensionless group velocity dispersion (GVD), and $V(\psi)$ is a nonlinear term. For the case $\beta_2 = 0$, Eq. (S2) drives transverse displacement of purely spatial optical vortices through the term $\nabla_\perp^2 \psi$, as shown in refs. (7,8). Here, the additional term containing $\beta_2$ drives diffraction in the local space ($\xi$) direction. Therefore, as the beam propagates, the position of the vortex ring moves temporally as well.

Suppose at $z_0$ the position of the vortex ring is $\mathbf{r}_{vortex} = (\xi_0, r_0)$, and at $z_0 + dz$ the position of the vortex ring is $\mathbf{r}_{vortex} + d\mathbf{r}_{vortex} = (\xi_0 + d\xi, r_0 + dr)$. Then
$$\psi(r_0 + dr, \xi_0 + d\xi, z_0 + dz) \approx \psi(r_0, \xi_0, z_0) + \frac{\partial \psi}{\partial r} dr + \frac{\partial \psi}{\partial \xi} d\xi + \frac{\partial \psi}{\partial z} dz.$$

But since $\psi = 0$ at the vortex, this leads to

$$\nabla_{ST} \psi \cdot d\mathbf{r}_{vortex} = -\frac{\partial \psi}{\partial z} dz \quad, \tag{S3}$$

where the spacetime gradient is defined as $\nabla_{ST} = \frac{\partial}{\partial r}\hat{\mathbf{r}} + \frac{\partial}{\partial \xi}\hat{\xi}$.

Following the method of ref. (7) as applied to OAM vortices, we approximate the local form of the STOV as a spacetime "$R$-vortex" $\psi_{vortex} \equiv (\xi - \xi_0) \pm i(r - r_0)$ of charge $\pm 1$ with a linear phase winding about $(\xi_0, r_0)$, embedded in a background field envelope $\psi_{bg}$ such that $\psi = \psi_{bg} \psi_{vortex}$. If we take $\psi_{bg} = \rho e^{i\chi}$, where $\rho$ and $\chi$ are the real amplitude and phase of the background field, then substitution of $\psi = \psi_{bg} \psi_{vortex}$ into Eq. (S2) yields

$$k \frac{\partial \mathbf{r}_{vortex}}{\partial z} = \pm \frac{1}{\rho}\left(\hat{\mathbf{r}} \frac{\partial \rho}{\partial r} - \hat{\xi}\beta_2 \frac{\partial \rho}{\partial \xi}\right) \times \hat{\phi} + \left(\hat{\mathbf{r}}\frac{\partial \chi}{\partial r} - \hat{\xi}\beta_2 \frac{\partial \chi}{\partial \xi}\right) \pm \hat{\xi}\frac{1}{2r} \quad. \tag{S4}$$

This is Eq. (2) in the main text. As the pulse propagates along $z$, Eq. (S4) gives the next move of the STOV based on the current self-consistent background field $\psi_{bg}$.

STOV in spatio-spectral space

The STOV is a vortex in the spatio-temporal wavefront, but the experiment measures the spatial phase of the filamenting pulse by interference with a reference pulse that is centered at 800 nm. What is the signature of a STOV in spatio-spectral space?

Consider an '$r$' vortex ring centered at $(r_0, \xi_0)$ embedded in a Gaussian background field with a possible temporal chirp,

$$E(r, \xi) = E_0 e^{-r^2/w_r^2 - \xi^2/w_\xi^2} e^{-ia\xi^2} [\xi - \xi_0 \pm i(r - r_0)], \tag{S5}$$

where $w_r$ and $w_\xi$ are transverse and longitudinal widths, respectively, and $a$ is a chirp parameter. The Fourier transform along the $\xi$ axis is, up to a complex coefficient,

$$E(r,\omega_\xi) \propto e^{-r^2/w_r^2 - \omega_\xi^2/(ia-w_\xi^{-2})} \left\{ (2+2iaw_\xi^2)\xi_0 - iw_\xi^2\omega_\xi \pm 2(r-r_0)(aw_\xi^2 - i) \right\}.$$

Here we have defined a spectrum-like quantity $\omega_\xi = \omega/v_g$. In spatio-spectral space the vortex is centered at

$$r_0' = r_0 \mp \frac{\xi_0}{aw_\xi^2}, \omega_{\xi,0} = 2\xi_0 \left( \frac{1}{aw_\xi^4} + a \right). \tag{S6}$$

If the chirp parameter were zero ($a = 0$), the vortex ring would be located at infinity. However, for a very small non-zero chirp parameter, the vortex is located inside the region where the field is non-zero. For example, for a 40 fs pulse, the group delay dispersion (GDD) $(1/2)\beta_2 = a/[8(a^2 + w_\xi^{-4})]$ only needs to be ~4 fs$^2$ in order for the vortex to overlap with the pulse in space and spectrum. In addition, for even small values of GDD, $r_0' \approx r_0$.

Energy current density

Without loss of generality, we take $\psi = ue^{i\Phi}$, where $u$ and $\Phi$ are the real amplitude and phase of the field. Based on conservation of energy we can derive an equation of the form $\frac{\partial}{\partial z}u^2 = -\nabla \cdot \mathbf{j}$ where $\mathbf{j}$ is the energy current density in the laser pulse frame and $u^2 = |\psi|^2$ is the *normalized energy density*. Following an argument similar to (9), in order to derive $\mathbf{j}$ we start with $\frac{\partial}{\partial z}u^2 = \frac{\partial}{\partial z}(\psi^*\psi) = \psi^*\frac{\partial}{\partial z}\psi + c.c.$ and replace $\frac{\partial}{\partial z}\psi$ with (S2). Near the vortex, the amplitude of the field is close to zero, so loss mechanisms like ionization and molecular rotations can be neglected, resulting in $\psi^*\left(\frac{1}{2ik}\right)V\psi + c.c. = 0$ ($V\{\psi\}$ is a real quantity). Furthermore, for other terms, after cancelling out the purely imaginary terms with their complex conjugates, we obtain $\frac{-1}{2ik}\psi^*\nabla_\perp^2\psi + c.c = -\frac{1}{k}(u^2\nabla_\perp^2\Phi + \nabla_\perp u^2 \cdot \nabla_\perp \Phi)$ and $\frac{\beta_2}{2ik}\psi^*\frac{\partial^2}{\partial \xi^2}\psi + c.c = \frac{\beta_2}{k}\left( u^2\frac{\partial^2}{\partial \xi^2}\Phi + \frac{\partial}{\partial \xi}u^2\frac{\partial}{\partial \xi}\Phi \right)$, resulting in an energy current density of

$$\mathbf{j} = \frac{1}{k}u^2(\nabla_\perp\Phi - \beta_2\frac{\partial\Phi}{\partial\xi}\hat{\xi}) \tag{S7}$$

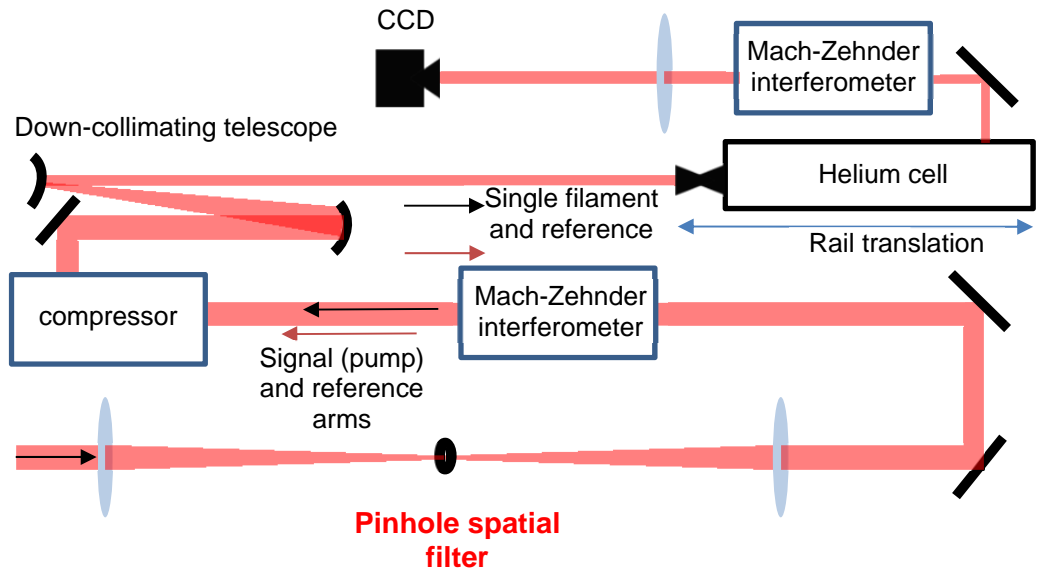

**Fig. S1.**

Experimental setup for measuring the in-flight intensity and spatial phase profiles of collapsing and filamenting femtosecond pulses over a ~2 m range.

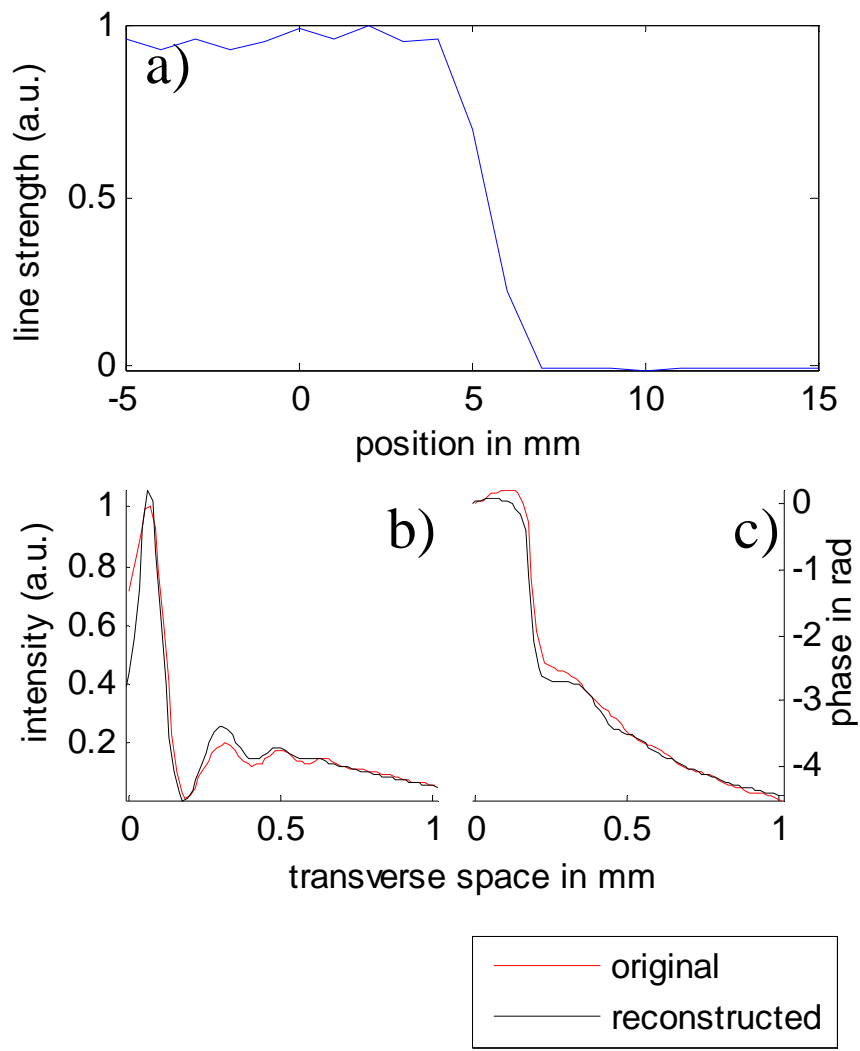

**Fig. S2**

(a) Line emission of the neutral helium $3\ ^3D - 2\ ^3P$, $\lambda=587.6$ nm transition from a tightly focused 800 nm pulse as a function of helium cell position. (b) Simulated filament intensity and (c) spatial phase at 800 nm (black) just before the ~4 mm air-helium transition, as well as reconstructed intensity and phase (red). The reconstruction is performed by propagating the solution 4 mm through the transition region followed by 50 cm of helium. This simulated far field is then back-propagated through 50.4 cm of vacuum to simulate reconstruction from imaging optics.

**Movie S1**

See http://lasermatter.umd.edu

**STOV birth and collision:** Initial pulse is collimated with Gaussian beam waist $w_0$= 1.3mm, 45 fs FWHM intensity beam, and energy 2.8 mJ. The beam is propagating in air. From z = 130cm to z = 160cm, the beam undergoes optical collapse, and then generates 4 STOVs from z = 163cm to z = 167cm. There is a STOV annihilation event at z = 204cm.

**STOV settles with main light bullet:** Same simulation input parameters as "STOV birth and collision", but with window zoomed in on the temporally foremost light bullet. We see the +1 STOV which forms with the pulse at optical collapse (z = 167 cm) settle around the light bullet. There is an additional vortex pair that forms at z = 219cm. The +1 STOV subsequently annihilates on axis at z = 232cm, while the -1 STOV is annihilated by another vortex that forms on axis at z = 243cm and is then almost immediately annihilated by another collision at z = 244cm.

**References**

1 . M. Takeda, H. Ina, and S. Kobayashi, J. Opt. Soc. Am. 72, 156 (1982)
2 . T. R. Clark and H. M. Milchberg, Phys. Rev. Lett. **78**, 2373 (1997)
3 . A. Ting *et al*., Applied Optics **44**, 1474 (2005)
4 . J. P. Palastro, T. M. Antonsen Jr., and H. M. Milchberg, Phys. Rev. A *86*, 033834 (2012)
5 . A. Couairon and A. Mysyrowicz, Phys. Rep. **441**, 47 (2007)
6 . S. Skupin *et al*., Phys. Rev. E **74**, 056604 (2006)
7 . D. Rozas, C. T. Law, and G. A. Swartzlander, Jr., J. Opt. Soc. Am. B **14**, 3054 (1997)
8 . G. Indebetouw, J. Mod. Opt. **40**, 73 (1993)
9 . D. Faccio *et al*. Opt. Expr. 17 8193 (2009)